\documentclass[twocolumn, showpacs, superscriptaddress,amsmath, amsfonts,prb]{revtex4}
\pdfoutput=1
\usepackage{graphicx}

\begin{document}

\title{A  thermodynamic theory of filamentary resistive switching}
\author{V.G.Karpov}
\affiliation{Department of Physics and Astronomy, University of Toledo, Toledo OH 43606, USA}
\author{D. Niraula}\affiliation{Department of Physics and Astronomy, University of Toledo, Toledo OH 43606, USA}
\author{I. V. Karpov}
\affiliation{Components Research, Intel Corporation, Hillsboro, Oregon 97124, USA}
\author{R. Kotlyar}
\affiliation{Process Technology Modeling, Intel Corporation, Hillsboro, Oregon 97124, USA}

\begin{abstract}
We present a phenomenological theory of filamentary resistive random access memory (RRAM) describing the commonly observed features of their current-voltage characteristics. Our approach follows the approach of thermodynamic  theory  developed earlier for chalcogenide memory and threshold switches and largely independent of their microscopic details. It explains, without adjustable parameters, such features as the domains of filament formation and switching, voltage independent current in SET and current independent voltage in RESET regimes, the relation between the set and reset voltages, filament resistance independent of its length, etc. Furthermore, it expresses the observed features through the material and circuitry parameters thus paving a way to device improvements.
\end{abstract}
\maketitle
\section{Introduction: questions}\label{sec:intro}
Filamentary resistive random access memory (RRAM) devices have been a subject of intensive investigations for more than a decade. In spite of a significant amount of data accumulated for various materials systems, many aspects of device operations are not understood and, unlike e. g. spin transfer torque memory (STTM), \cite{slonczewski1996} their understanding remains rather limited, and a sufficient theory of the resistive switching phenomena is not yet available.

\begin{table}[h!]
\caption{Outstanding questions about RRAM current-voltage (I/U) characteristics }
  \begin{tabular}{|p{4.7cm} | p{3.7cm} |}
    \hline
   Domain marked in Fig. \ref{fig:cartoon}\footnotemark[1]$^,$\footnotemark[2]& Question \\ \hline
   A-B, Switching at threshold voltage $U_T$.    & What is the nature of the snapback at $U_T$ and its dependence on material and circuit parameters, temperature and voltage rate?  \\  \hline
   B-C, Vertical I/U at $U_{\rm SET}$.  & $U_{\rm SET}$ vs. material parameters.  \\  \hline
   C-0-D, The ON state (low resistance) domain & Filament radius and resistance $R\propto 1/I_{\rm SET}$ vs. material parameters. \\  \hline
   D-E, Switching to $R_{\rm RESET}$ at $U_{\rm RESET}$  & Why does the resistance increase past $U_{\rm RESET}$? What is the nature of snapforward? Expression for $U_{\rm RESET}$, $I_{\rm RESET}$, and snapforward ratio $I_D/I_E$.\\  \hline
   E-F, The current saturating to $I_{\rm R,SAT}$ or slightly increasing towards $U_{\rm STOP}$ (`horizontal I/U') & Expressions for $I_{\rm R,SAT}$ and/or $I(U)$ in E-F.  \\  \hline
  \end{tabular}\label{tab:questions}
  \footnotetext[1]{Here we limit ourselves to the case of bipolar RRAM undergone the filament forming process.}
  \footnotetext[2]{We do not discuss here the OFF state (high resistance) domain F0A, in which IV characteristics are determined by the insulating material properties without structural transformations.}
  \end{table}
There is a consensus about the crucial role of conductive filaments (CF) determining RRAM operations. CF can break switching the device into RESET state. Reestablishing CF would switch the system into its SET state. The existing models of CF are either qualitative or entirely numerical, containing a number of adjustable parameters.

\begin{figure}[t]
\includegraphics[width=0.40\textwidth]{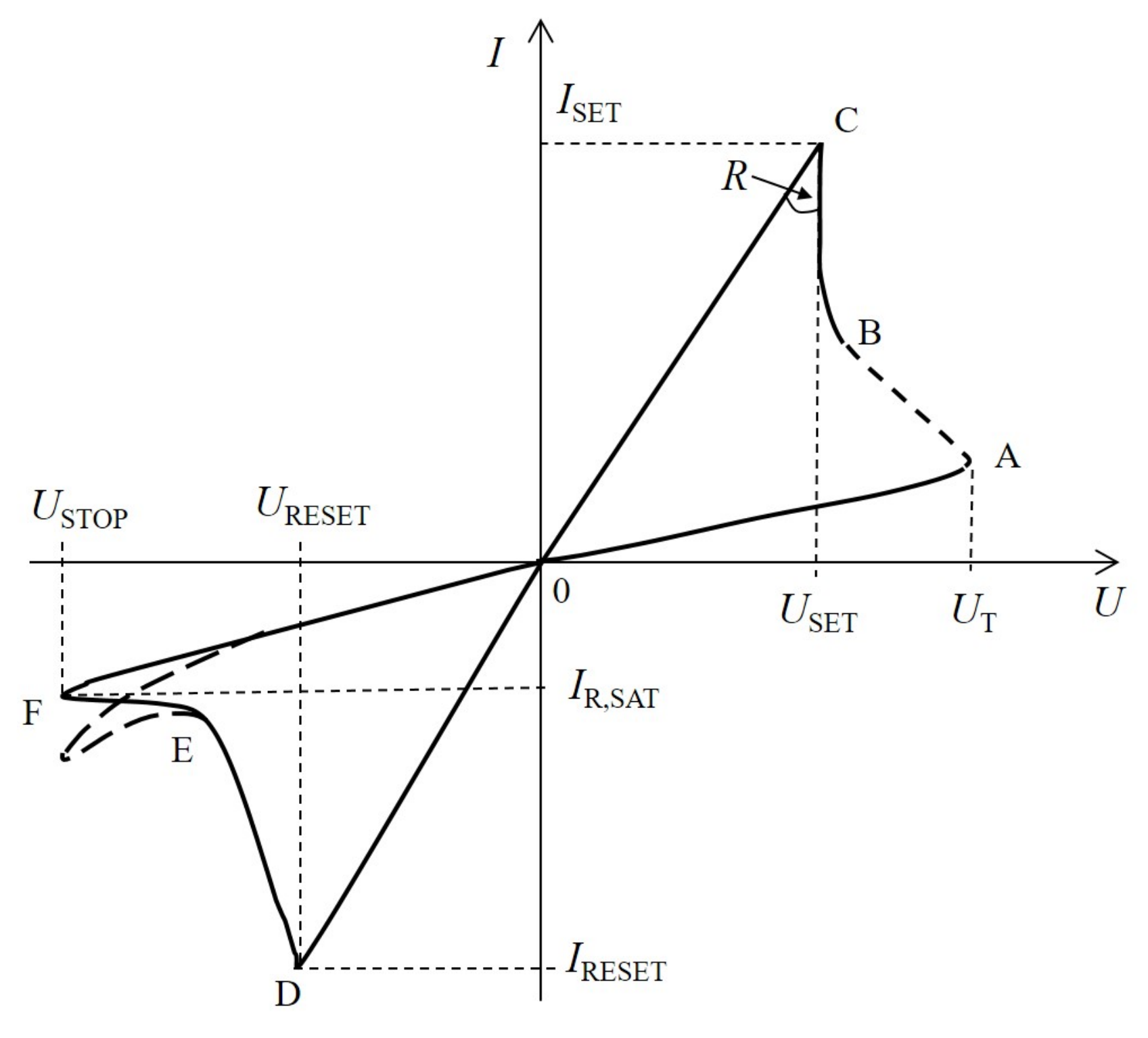}
\caption{A sketch of the typical current voltage characteristics \cite{fantini2012,wouters2012,wei2015,sawa2005,beck2000} showing various domains in Table \ref{tab:questions}. $U$ is the voltage across the device different from the power source voltage $V$ due to the series load resistor; $U_{\rm STOP}$ is determined by the maximum absolute value of voltage $V$ during the reset process. The slope $R$ at point C is the conductive filament resistance. Note that in experiments, the SET current $I_{\rm SET}$ is defined as the maximum (compliance) current allowed. The dashed fragment in the third quadrant shows the sometime observed deviations from the current saturation behavior with $|I_F|\gtrsim |I_E|$. The dashed fragment in the first quadrant represents the switching part of the SET process where the data points between A and B may not be measurable.\label{fig:cartoon}}.
\end{figure}
This work introduces a quantitative phenomenological  theory of RRAM answering several  outstanding questions.
They are listed in Table \ref{tab:questions} referring to the sketch of typical current-voltage characteristics in Fig. \ref{fig:cartoon}.
As an example, we elucidate the symmetry $U_{\rm RESET}=-U_{\rm SET}$, similarity between $U_{\rm RESET}$ and $U_{\rm SET}$ in various systems, vertical and horizontal domains of the current-voltage characteristics in the SET and RESET regions respectively, etc.

Our theory below provides quantitative answers to the questions of Table \ref{tab:questions} in the framework of a phenomenological analysis that does not specify the microscopic structure of CF or the details of chemical composition. Instead, it concentrates on generic thermodynamic properties consistent with the data. This is achieved by introducing the chemical potentials of different phase states involved and considering the system free energy that includes the thermal, the electric, and the chemical components.

The consideration is organized as follows. In Sec. \ref{sec:mod} we introduce our model of CF with a previously overlooked property of the polarity dependent electric charging. Sec. \ref{sec:proc} describes the thermodynamic analyses of nucleation and growth process related to the domains in Fig. \ref{fig:cartoon} and Table \ref{tab:questions}. Sec. \ref{sec:quant} presents our quantitative results. The conclusions are given in Sec. \ref{sec:concl}.

\section{Filament model}\label{sec:mod}

Any CF model has to address a significant fact that CF resistance is practically independent of its length $h$.  An often assumed picture of CF  postulates a local geometrical constriction responsible for CF resistance (hourglass model \cite{degrave2014,kim2010}). In that model, the RESET and SET processes are attributed to the destruction and restoration of the constriction.  Experimentally, it was found that CF can have a truncated cone shape. However the radii of that cone faces are typically different by only a numerical factor, \cite{celano2014,privetera2013} $r_1/r_2\sim 2$, not  significant enough to attribute the entire cone resistance to its narrow region [the truncated cone resistance, \cite{halliday} $R_{\rm cone}= \rho h/(\pi r_1r_2)$ where $\rho$ is the resistivity].


\begin{figure}[thb]
\includegraphics[width=0.35\textwidth]{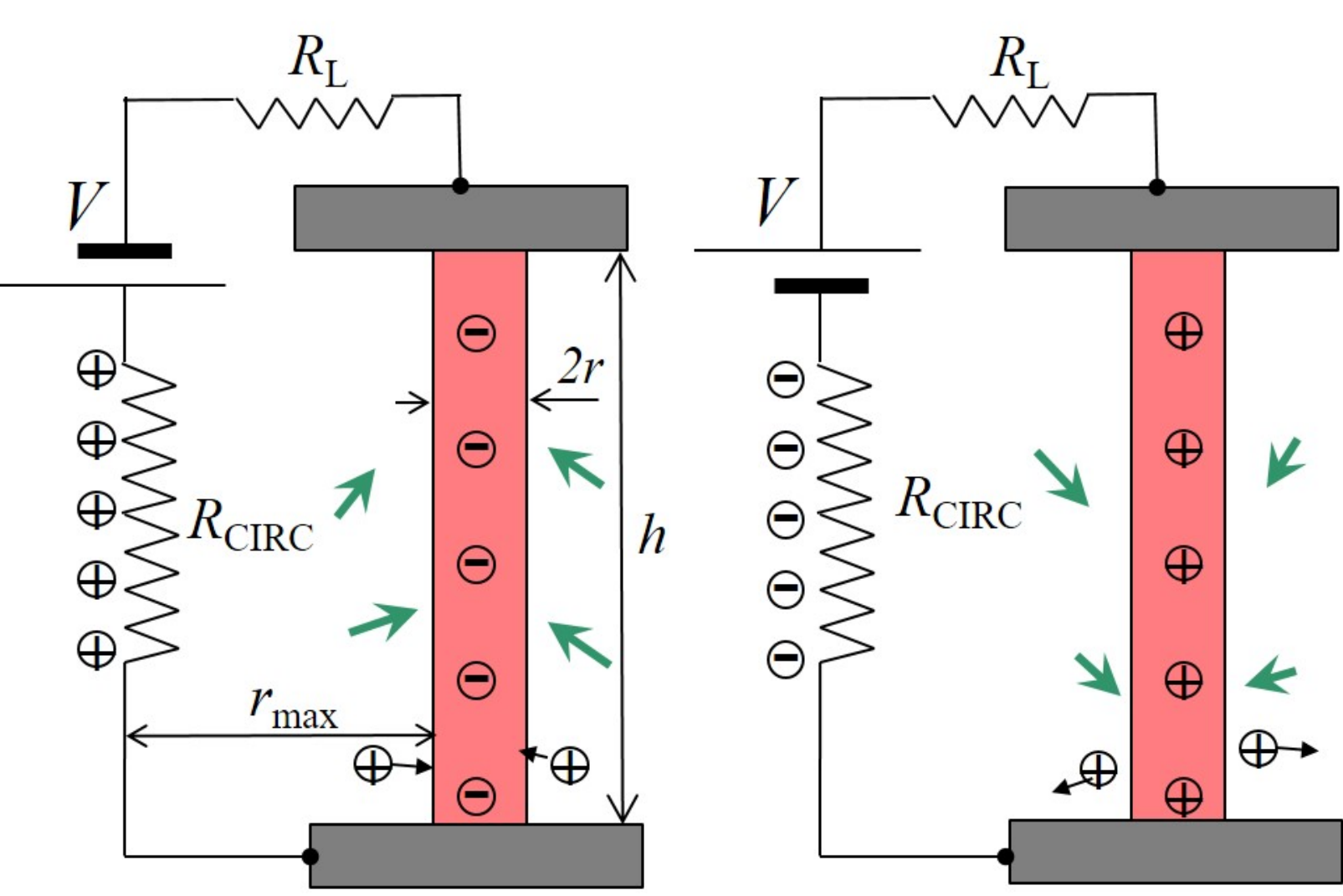}
\caption{A conductive filament under biases of different polarities accumulates the electric charges (denoted by  $\ominus$ and $\oplus$) creating the electric field that attracts (left) or repels (right) ions (for specificity shown as $\oplus$ outside the filament). $R_{\rm CIRC}$ represents the system wires with their own capacitance and charges, characterized by linear dimension $r_{\rm max}$. $R_L$ is the load resistance; $R_L\gg R_{\rm CIRC}$. Short arrows represent the electric polarization along the energetically favorable directions parallel to the local field. \label{fig:filam}}
\end{figure}

Our phenomenological theory below does not specify CF structure treating it as a formed conductive cylinder; we will not assume that its conical shape is essential. The CF resistance independent of its length $h$ will be explained without the assumption of its determining constriction [see Eqs. (\ref{eq:param1}) and (\ref{eq:USET1}) below]. As a novel structure-independent feature,  our model in Fig. \ref{fig:filam} includes the electric charges accumulated by CF due to its electric capacitance as explained next; they induce electric polarization that couples the host and CF.

The latter polarization can be caused by either redistribution of ions or  reorientation of local ferroelectric domains. The existence of such domains does not necessarily imply that the host material exhibits bulk ferroelectricity. It is known indeed that making HfO$_2$ ferroelectric requires particular doping and growth techniques stabilizing the nonequilibrium orthorhombic phase.\cite{muller2015} However, the microscopic ferroelectric domains can exist locally due to the stress or field related conditions around CF. These phenomena were observed even for amorphous  morphologies. \cite{xu1999} For a particular case of Hf based RRAM, it should be noted that the high dielectric permittivity $\varepsilon\approx 25$ is mostly due to ionic displacements \cite{zhao2002} [as follows from the comparison with the square of refraction index ($\approx 2.1$)] and can be pinned by some defects.

To describe CF charging we recall the well known model of two long parallel wires of radius $r$ each separated by distance $r_{\rm max}\gg r$ and connected to the power source through resistance $R_L\gg R$. The capacitance ($C$) and charge ($\beta$) per length, and the radial electric field at the wire surface ($E_r$) are (in Gaussian units),
\begin{equation}\label{eq:filpar}
C=\left[2\ln\left(\frac{r_{\rm max}}{r}\right)\right]^{-1},\quad\beta =IR_LC,\quad E_r=\frac{2\beta}{r} \end{equation}
where $I$ is the current. Because $r_{max}\gg r$, changing $r_{max}$ by a numerical factor or even by an order of magnitude, (say, from 1 mm to 1 cm) will not significantly change the results in Eq. (\ref{eq:filpar}), which are not sensitive to the circuitry design.

Furthermore, a realistic analysis takes into account that  CF charging simultaneously creates the corresponding image charges in metal electrodes, which screens the lateral electric field at distances of the order of $h$ from CF. \cite{cooray2007} As a result, $r_{\rm max}$ under the logarithm in Eq. (\ref{eq:filpar}) should be replaced with $h$, which makes it fully independent of a particular circuit design.

Eq. (\ref{eq:filpar}) formally predicts CF generated radial electric field that should disappear when the current is turned off, $I=0$. In Sec. \ref{sec:proc} below, we consider the atomic rearrangements (ion or ferro- displacement ) electric polarization ${\bf \Pi}$ caused by that field. Such a polarization possesses significant inertia making it long-lived   after the current is turned off. Furthermore, we will show that a self-consistent state of CF charge and surrounding polarization can form a polaron like stable or metastable state.

$C$ in Eq. (\ref{eq:filpar}) should not be mixed with the specific capacitance of a stand alone thin metal needle  analyzed since Maxwell, \cite{maxwell,mcdonald2003,scharstein2007} and given, per length, by
\begin{equation}\label{eq:selfcap} C_0=\left[2\ln\left(\frac{h}{r}\right)\right]^{-1},\end{equation}
numerically close to $C$. The difference between $C$ and $C_0$ effects is that the former accumulate charges due to electric current flow, which depends on the current polarity, while the latter acquires charges in response to the electric potential difference between the material of CF and its surrounding material, i.e. polarity independent. For example, estimating the Fermi energies difference between Hf dominated CF and its HfO$_2$ surrounding \cite{kruchihin2015} as $\delta E_F\sim 1-2$ eV, the current independent charge per length becomes,
\begin{equation}\label{eq:selfcharge}
\beta _0=\frac{C_0\delta E_F}{e}\end{equation}
where $e$ is the elemental charge. Depending on the relation between $\beta$, $\beta _0$, one can predict current $I$ driven changes in the electric field $E_r$ contributing to CF related operations in both bipolar and unipolar modes.

Note that the concept of CF charging is model independent . While the filament capacitance is numerically insignificant, say $C\sim 0.1$ pF/cm, its electric effect is strong due to a relatively small radius $r\ll h$.  The field in Eq. (\ref{eq:filpar}) is strong compared to the field $IR_L/h$ between the electrodes, with the ratio $E_r/E= (h/r)/\ln(h/r)\gg 1$. Through the radial electric field, the bias polarity will stimulate redox or other processes affecting CF size and morphology. In particular, the above predicted electric field $E_r$ could explain the radial drift of ions assumed by the ion drift models for RRAM operations (see e. g. Ref. \onlinecite{ielmini2011} and references therein). However, our phenomenological treatment here does not explicitly specify the underlying microscopic models.


Another model independent statement pertains the fact that CF does not undergo any significant changes (representing a long lived conductive channel) when the voltage across the device is between $U_{\rm SET}$ and $U_{\rm RESET}$, while voltages beyond that interval cause significant CF transformations. In particular $R=dV/dI$ determines the resistance of that long-lived CF that does not change between $U_{\rm SET}$ and $U_{\rm RESET}$, i. e.
\begin{equation}\label{eq:R=R}R(U_{\rm SET})=R(U_{\rm RESET}).\end{equation}

Phenomenologically the property of $U_{\rm SET}$ and $U_{\rm RESET}$ to confine the regime of CF stability, means that they play the role of `freezing/unfreezing' voltages, such that the temperature above $U_{\rm SET}$ and below $U_{\rm RESET}$ must be higher than some freezing temperature $T_f$, while it is below $T_f$ when $U_{\rm RESET}<U<U_{\rm SET}$. $T_f$ can correspond to a particular phase transition, such as e. g. glass transition, \cite{moynihan1974} but also, in general, to any thermally activated process.   Assuming the activated atomic transformation with the characteristic time $\tau _0\exp(-W_a/kT)$  where $\tau _0\sim 10^{-13}$ s is the characteristic period of atomic vibrations, $k$ is the Boltzmann's constant, and $W_a$ is the activation energy, $T_f$ is determined by the condition (similar to Refs. \onlinecite{moynihan1974,ald1990,street1991}),
\begin{equation}\label{eq:Tf} \tau =\tau_0\exp(W_a/kT_f)\end{equation}
where $\tau$ represents the voltage pulse width or $[d(\ln U)/dt]^{-1}$ for a continuously varying voltage $U(t)$.

The Joule heat related temperature change is described by, $\delta T=T_f-T_0=\tau _T P/kN_a$ with $\tau _T$ being the thermalization time, $T_0$ the room temperature, $P=U^2/R$, and $N_a$ is the number of degrees of freedom (roughly equal the number of atoms) in the region involved. Therefore, the freezing/unfeezing condition takes the form
\begin{equation}\label{eq:rate}
\tau _T\frac{U^2}{RN_a}= \frac{W_a}{\ln(\tau /\tau _0)}-kT_0.\end{equation}

Since the criterion in Eq. (\ref{eq:rate}) is satisfied for voltages $U_{\rm SET}$ and $R(U_{\rm SET})=R(U_{\rm RESET})$ according to Eq. (\ref{eq:R=R}), we conclude that it is satisfied when $U_{\rm RESET}=-U_{\rm SET}$ thus elucidating the latter relation pointed among the outstanding challenges in Table \ref{tab:questions}. We note that our model does not rely on details of any particular microscopic mechanism for Eq. (\ref{eq:rate}) unlike, say, Eqs. (1) - (6) in Ref. \onlinecite{ielmini2011}. In particular, it remains applicable to the processes in glasses of phase change memory where bipolar switching was recently observed. \cite{coiocchini2016}

Furthermore, Eq. (\ref{eq:rate}) predicts that $U_{\rm SET}$ and $U_{\rm RESET}$ will change logarithmically with $\tau$, which was observed. \cite{fantini2012,maestro2015,schindler2009,yalon2015a} We note that in a noncrystalline system, the activation energies generally vary between different local regions in a manner described in Sec. \ref{sec:amorph} below.

\section{Chemical potentials}\label{sec:proc}

\subsection{Three states of the system}\label{sec:three}
Similar to the standard phase transitions, we assume CF transforming through the nucleation and growth stages. The newly nucleated phases may not be immediately stable or even long lived. We consider a possibility that they initially appear  as unstable, having to undergo further transformations towards stability. It was independently argued indeed \cite{erdemir2009} that in polymorphic systems, nucleation can evolve in two steps, through an intermediate metastable phase. Also, it has been experimentally observed that CF can be annealed at high enough temperature, \cite{govoreanu2013,park2016} i. e. it presents a metastable state  lived long enough  to have practical significance as a nonvolatile memory.

\begin{table}[t!]
\caption{Processes and chemical potentials corresponding to different domains in Fig. \ref{fig:cartoon}}
  \begin{tabular}{|c |p{5cm} | p{2.1cm} |}
    \hline

   Domain & $\quad\quad\quad\quad$ Process & CP\footnotemark[1] \\ \hline
   A-B  & nucleation and longitudinal growth of a narrow unstable CF shorting  between the electrodes & $\mu _{uc}=\mu _i+\delta\mu _1>\mu _i$ \\  \hline
   B-C  & radial growth of the long lived charged CF and its stabilizing polarization near point C making CF long lived & $\mu _{uc}\rightarrow \mu _{mc}=\mu _{uc}-\delta\mu _2 <\mu _{uc}$, $\quad\quad$ $\mu _{mc}>\mu _i$ \\  \hline
   C-0-D  & long lived metastable CF changing the charge polarity at point O & $\mu _{mc}$  \\  \hline
   D-E  & unfreezing oppositely charged metastable CF in the `wrong polarization' environment, CF break up via nucleation of insulating gap  & $\mu _{mc}\rightarrow \mu _{i}$ \\  \hline
   E-F  & increase in the insulating  gap to its steady state width  & $\mu _{i}$ \\  \hline

  \end{tabular}\label{tab:processes}
  \footnotetext[1]{CP stands for the chemical potentials of the insulating phase ($\mu _i$), unstable CF ($\mu _{uc}$), and metastable CF ($\mu _{mc}$) phases illustrated in Fig. \ref{fig:CP}.}
\end{table}

\begin{figure}[h]
\includegraphics[width=0.5\textwidth]{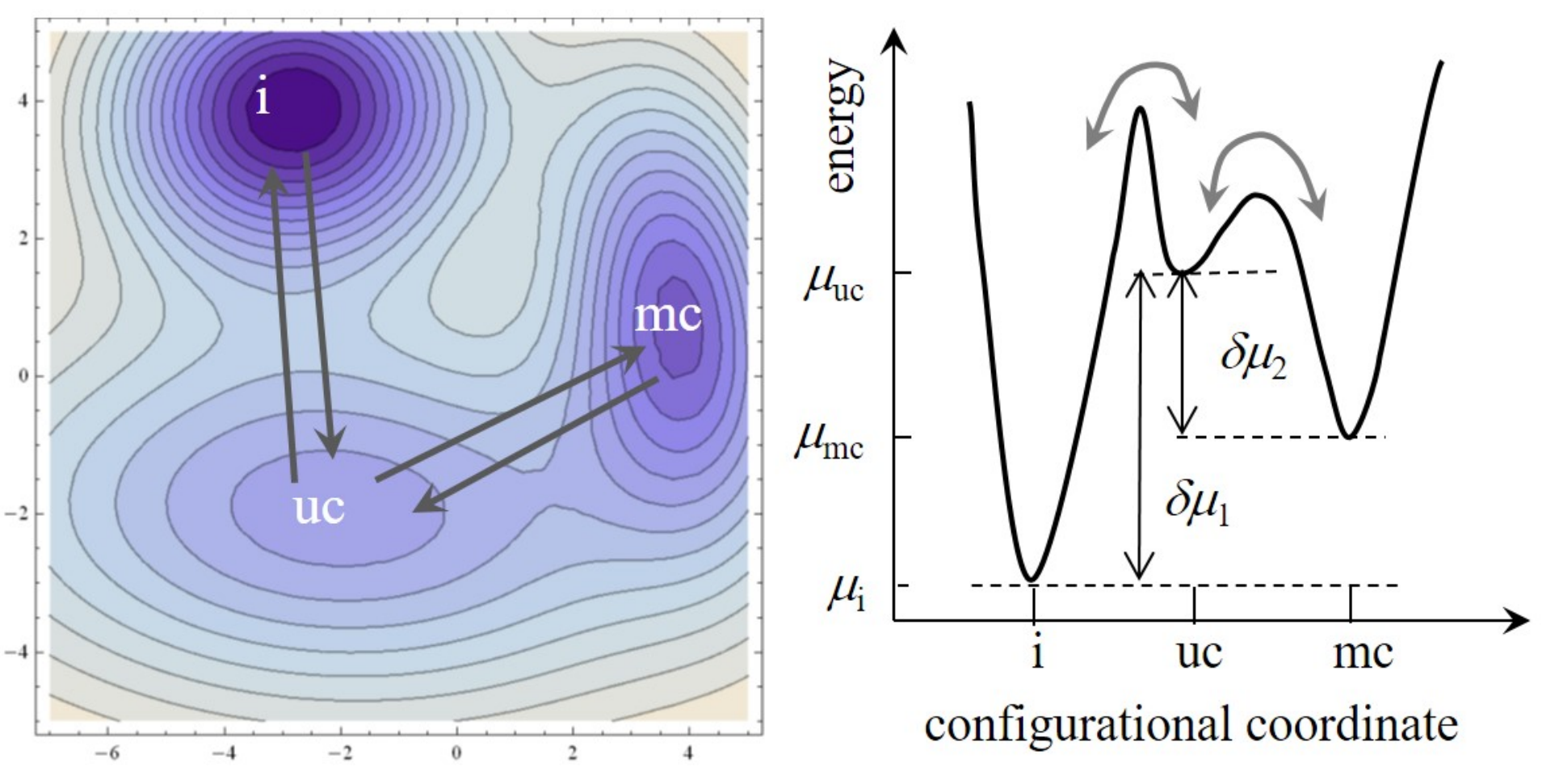}
\caption{Left: A contour plot of the system chemical potentials in 2D space of unspecified configurational coordinates showing three distinct minima corresponding to the insulating (i), unstable conductive (uc), and metastable conductive (mc) phases and their related barriers.  Right: 1D presentation of the same along an unspecified coordinate.  Arrows represent transformations between mc- and uc-, and uc- and i- phases where the energy barriers are relatively low.    \label{fig:CP}}
\end{figure}
\begin{figure}[h!]
\includegraphics[width=0.47\textwidth]{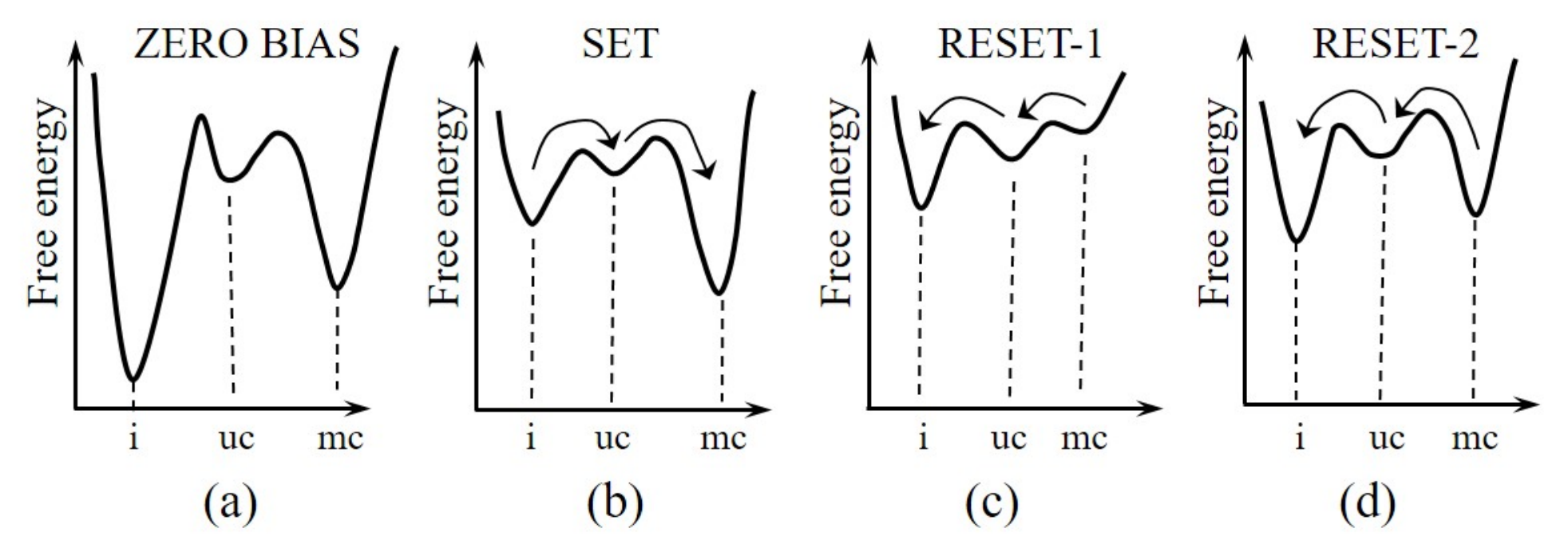}
\caption{Free energy vs. configurational coordinate under different electric biases corresponding to Fig. \ref{fig:cartoon}. (a) Zero bias, (same shape as in Fig. \ref{fig:CP}); (b) Point A, bias $U_T$ triggering SET process by nucleation and radial growth of CF; (c) point D, the opposite polarity charge triggering RESET via CF breakdown by nucleation of the insulating gap; (d) point E, strong field and composite CF combining an insulating gap and conductive domains in the final stage of RESET. See more explanations in the text.    \label{fig:CPF}}
\end{figure}
Table \ref{tab:processes} specifies our model processes and their corresponding chemical potentials related to various domains in Fig. \ref{fig:cartoon}. The field induced nucleation at the threshold voltage $U_T$  is followed by the longitudinal growth of a narrow CF that is unstable without the electric field. As shown below, its subsequent radial growth is characterized by resistance $R$ inversely proportional to the current; hence, the vertical current voltage characteristic at $U_{\rm SET}$. The chemical potential of the structure constituting that unstable CF, is higher than that of the insulating host, $\mu _{uc}>\mu _i$, as reflected in Table \ref{tab:processes}.

The relations between the chemical potentials of insulating, unstable (short-lived) conductive, and metastable (long-lived) conductive phases and their corresponding thermodynamic barriers are illustrated in Fig. \ref{fig:CP} for zero electric field, and in Fig. {\ref{fig:CPF} for finite electric biases in RRAM operation. The barriers describe energetically unfavorable configurations through which the system evolves towards a metastable or stable state.


While energetically most favorable under zero bias [Fig. \ref{fig:CPF} (a)], the insulating phase significantly increases its energy ($\propto E^2$) under electric bias due to the strong electric field  $E=U/h$. Assuming as usual a load resistance $R_L$ in series with the device resistance $R$, the source voltage $V$ corresponds to the device voltage $U=VR/(R+R_L)$ considerably lowered by CF that introduces a low resistance $R$ path. Therefore, under electric bias, the conductive states have lower energy than the insulating one as reflected in Fig. \ref{fig:CPF} (b). In that same diagram arrows show the processes of nucleation and growth through a short-lived state (uc) with the left barrier playing the role of nucleation barrier decreased by the field as described in Sec. \ref{sec:thresh} below. The latter short-lived state then decays into the long-lived conductive state (mc) that is lower in energy than the insulating state (i); this sequence constitutes the SET process.

Presented in Fig. \ref{fig:CPF} (c) is the system with SET formed CF under the instantaneously reverted bias polarity. The inherited polarization conflicting with the instantaneously acquired CF opposite charge strongly increases the free energy of a formerly stabilized CF state making it unstable and triggering CF breakdown by forming an insulating gap. This constitutes the first stage of the RESET process described more quantitatively in Sec. \ref{sec:insgap} next.

Shortly after CF polarity reversal, the surrounding polarization will realign correspondingly lowering CF energy as depicted in Fig. \ref{fig:CPF} (d). The subsequent growth of the insulating gap will proceed through the energetically unfavorable short-lived (uc) state presented by arrows in Fig. \ref{fig:CPF} (d). This constitutes the second stage of RESET quantitatively described in Sec. \ref{sec:gapgr} below.

%
%

Phenomenologically, $\delta \mu _1=\mu _{uc}-\mu _i$ remains a material parameter. It can be estimated for specific CF models, such as formed by oxygen vacancies in HfO$_2$. Assuming their relative concentrations in the bulk and CF to be respectively  $n_{b}\sim 0.1$ and $n_{\rm CF}\sim 1$ and using the results for dilute solutions ($n_{b}\ll 1$), \cite{landau1980} one gets
\begin{equation}\label{eq:mu1}\delta \mu _1=\frac{kT}{a_0^3}\ln(n_{\rm CF}/n_b)\end{equation}
where $a_0^3$ is the volume per vacancy, roughly equal the atomic volume. Based on the temperature measurements, \cite{yalon2015} we take $T\sim 600$ K. Taking also $a_0\sim 0.2$ nm yields $\delta\mu _1\sim 10^9$ J/m$^3$.

At a certain radius and resistance satisfying the criterion in Eq. (\ref{eq:rate}) with $U=U_{\rm SET}$, CF becomes stabilized by the host polarization, as explained in Sec. \ref{sec:mod} before the paragraph containing Eq. (\ref{eq:filpar}). Its structure remains frozen in the interval of voltages $U_{\rm RESET}<U<U_{\rm SET}$. The polarization ${\bf \Pi}$ contribution to the chemical potential is given by, \cite{landau1984} $\delta\mu _2={\bf \Pi\cdot E}$ where ${\bf E}$ is the electric field, which is due to the charged CF for the case under consideration.

We describe the polarization assuming that it significantly screens the filament field, i. e., ${\bf E}\approx -4\pi {\bf \Pi}$, and
\begin{equation}\label{eq:mu2}\delta \mu _2=|E\Pi |\approx 4\pi \Pi ^2.\end{equation}
According to definition, the polarization $\Pi =(ea)n$ where $ea$ is the elemental dipole corresponding to the elemental (ion) charge $e$ shifted over distance $a$, and $n$ is the concentration of such dipoles. We take the typical $a\sim 0.1$ nm, and $n\sim 10^{22}$ cm$^{-3}$, which yields $\delta\mu _2\sim 10^{9}$ J/m$^3$. In spite of the order of magnitude coincidence, $\delta\mu _1\sim \delta\mu _2$, one should assume $\delta\mu _1>\delta\mu _2$, on empirical grounds reflected in Table \ref{tab:processes}. Note that the above estimated polarization does not require significant diffusion of ions in the host material and thus can be fast enough to explain the observed fast transformations.

%

With voltage $U$ across the device changing its polarity, so does the electric charge density $\beta$ on CF, and its corresponding electric field ${\bf E}$. Therefore, the former polarization becomes energetically unfavorable, leading to the chemical potential $\mu _{mc}+|{\bf \Pi\cdot E}|>\mu _{mc}$ and triggering CF disruption at $U_{\rm RESET}$, when the criterion in Eq. (\ref{eq:rate}) is satisfied and CF structure thaws off. The disruption creates an insulating gap, which can grow further as described in Sec. \ref{sec:insgap} and \ref{sec:gapgr} below.

Finally, we note that a three-phase model similar to that of Fig. \ref{fig:CP} can be developed for the alternative case when the chemical potential of conductive phase is the lowest, as, for example, takes place in the phase change memory structures. It was observed indeed that the structural transformations in phase change memory involve more than just two phases; \cite{meister2011} hence, a three-phase description relevant.

\subsection{Bound states of  CF charge and polarization}\label{sec:polar}
In connection with the concept of polarization stabilized CF, we would like to point at the possibility of the polaron-like bound states retaining the  CF charges even after the current $I$ [in Eq. (\ref{eq:filpar})] is turned off.  Indeed based on the standard thermodynamics of dielectrics, the polarization energy gain can be represented as \cite{landau1984}
\begin{equation}\label{eq:polen}\delta F=-\frac{h}{2}\int _r^h\Pi E_r(r') 2\pi r'dr'=(1-\varepsilon )h\beta ^2\ln\left(\frac{h}{r}\right).\end{equation} where we have taken into account that  $\Pi =E_r(\varepsilon -1)/4\pi$. 

One can analyze the possibility of persistent CF charging by adding to Eq. (\ref{eq:polen}) the energy loss terms $\beta hV$ and $(\beta h)^2/2(Ch)$ with $\beta$ and $C$ from Eq. (\ref{eq:filpar}). They present respectively the work done to move the electric charge through the voltage source $V$ and to charge the CF capacitor. Approximating $V\approx IR_L$, it is straightforward to see that persistent CF charging is energetically favorable if $\varepsilon >2+\ln (h/r)$. The latter condition takes place for high dielectric permittivity materials. More realistic estimates should include the polarization related  anisotropy, strains, and nonlinearity.

\subsection{Role of amorphycity}\label{sec:amorph}
Here we discuss the role of  amorphycity of the material phases involved. It is well known from the physics of amorphous systems, that they are nonequilibrium gradually decreasing their energies with time (aging). In particular, (see Ref. \onlinecite{karpov2007} and references therein) the amorphous structure relaxation processes are responsible for the observed drift of parameters in phase change memory based on chalcogenide glasses.

The atomic configurations undergoing structural transformations are described as double well atomic potentials characterized by random thermodynamic barriers $W_B$. The probabilistic distribution of random barriers $W_B$ is approximated as uniform,
\begin{equation}
g(W_B)\approx 1/\Delta W_B, \quad \Delta
W_B=W_{B,max}-W_{B,min}\label{eq:prob}\end{equation} between the two
boundary values. That makes their relaxation times distribution reciprocal in $t$, and its related change in the system  energy is logarithmic in time, \cite{karpov2007}
\begin{equation}\label{eq:dmt}\delta\mu =\delta\mu _{min}+(\delta \mu _{max}-\delta\mu _{min})f(t), \end{equation}
where the distribution function of relaxation times, is given by,
\begin{equation}
f(t)=\frac{kT}{\Delta W_B}\ln\left(\frac{t}{\tau _{min}}\right),
\quad \tau _{min}<t<\tau _{max},
\label{eq:log}\end{equation}
and
\begin{equation}\tau _{max (min)}=\tau
_0\exp(W_{B,max(min)}/kT).\label{eq:tmax}\end{equation}
$f(t)$ saturates at $f_{max}\equiv
f(\tau _{max})=1$ for times $t>\tau _{max}$ and can describe a
remarkably broad time interval ranging from $\tau _{min}$ shorter
than one microsecond to, say, $\tau _{max}\sim 10^5$ s assuming
$\tau _0\sim 10^{-13}$ s (characteristic atomic vibration time)
and $W_{B,max}=1$ eV as a rough guide estimate.

According to Eqs. (\ref{eq:dmt}) and (\ref{eq:log}), any structural transformation in Table \ref{tab:processes} and Fig. \ref{fig:CP} involving one or more amorphous components, will exhibit long time relaxation behavior following logarithmic dependence. In some cases, those underlying logarithmic dependencies reveal themselves in other temporal forms entering results in exponents or other functions, such as, e. g. temporal drift of resistance \cite{karpov2007} given by,
\begin{equation}\label{eq:Rdrift}R(t)=R(0)\left(\frac{t}{\tau _{min}}\right)^\nu,\quad \nu=\frac{Du_0}{\Delta W_B}\end{equation}
where $D$ is the deformation potential and $u_0$ is the saturated value of the relative volume change (dilation), so that $\nu\sim 0.03$. The underlying mechanism is the material deformation changing the Fermi energy and resistance.

Long time logarithmic type relaxations in RRAM devices have been observed. \cite{fantini2012,maestro2015,schindler2009} Yet another evidence of random double-well atomic potentials is the 1/f noise (see details in Sec. 8.3.3 of Ref. \onlinecite{kogan1996}). 1/f noise corresponds to the self-correlation function (also known as the Pearson correlation coefficient) logarithmically decaying with time. \cite{hooge1997} Therefore, the recently observed \cite{fantini2015} correlation coefficient decaying linearly in $\log t$ for RRAM resistances measurements separated by time $t$, can be related to the above described random double well potentials.

The latter assertion requires a special comment explaining how the measurements in RRAM devices reveal mostly the random telegraph noises (RTN; see Ref. \onlinecite{maestro2016} and references therein)  rather the the 1/f noise. RTN are commonly related to double state fluctuators (double well potentials) when the number of such fluctuators is small. \cite{kogan1996} When the size of large systems with 1/f nose decreases to the degree that only a few  fluctuators left, then the noise acquire a behavior of RTN . Vice versa, the superposition of a great number of two state fluctuators corresponding to small devices with various relaxation times is seen as a 1/f noise. \cite{kirton1989,tsai2013}

The latter argument applies to RRAM filamentary devices where the effective volume contributing to operations is extremely small being limited to a fraction of CF undergoing structural transformations; similarly small is the number of contributing fluctuators corresponding to noises of not too low frequencies and revealing themselves via RTN signal. However, in extremely long time measurements, the number of significant fluctuators increases to include those with large relaxation times. Because a system with large number of fluctuators possesses 1/f noise behavior, it explains the observed logarithmic decay of correlation functions \cite{fantini2015}.

It is a general feature specific of the RRAM nano-sized devices that the number of double well potentials affecting CF is rather limited, i. e. not much larger than unity. Therefore, the results of reprogramming of a given device cannot be accurately described by averaging over continuous distribution of barriers characterizing the corresponding infinite system. This new situation of `nano-glass' remains to be further explored, although some important results are listed in Ref. \onlinecite{kogan1996}, Sec. 5.3. Here, we limit ourselves to stating that lack of self-averaging in a small system with random double well potentials leads to significant variations in their  created deformations,  \cite{karpov1992} and thus resistances, some of which will increase or decrease in the course of reprogramming. This type of behavior was observed with the magnitude of dispersion increasing towards small radius CF devices. \cite{fantini2012}

Also, we would like to point at the data on resistance variations as a function of the number of device reprogramming cycles $N$ [Ref. \onlinecite{chen2016}, Fig. 3(a)] exhibiting the dependence $R\propto N^{\nu}$ for both high and low resistance states. In the meantime, this or other specific device exhibits noticeable fluctuations between programming cycles.

We speculate that the latter dependence can be explained by Eq. (\ref{eq:Rdrift}) where $t$ is replaced with $N$. Such an interpretation implies  that increasing the number of reprogramming cycles increases the total time of exposure to elevated temperatures activating higher and higher barriers in the system. While this is not the standard temporal drift of parameters of a stand alone device, it can be described as the `reprogramming parameter drift'.

Finally, we would like to point at a difference between the temporal dependencies in Eq. (\ref{eq:dmt}) and that of Eq. (\ref{eq:rate}). The behavior in Eq. (\ref{eq:dmt}) is due to multiple random activation barriers in a broad interval of energies, characteristic of amorphous systems. To the contrary, Eq. (\ref{eq:rate}) describes a time dependence in a system with a single energy barrier $W_a$. In particular, it shows how a power of perturbation, necessary to change the material structure,  depends on the time during which it is exerted, while Eq. (\ref{eq:dmt}) predicts the long time relaxations independent of power injected.

\section{Quantitative analysis}\label{sec:quant}

\subsection{Nucleation events}\label{sec:nucl}
Here we consider the two nucleation events taking part among other processes listed in Table \ref{tab:processes}.
\subsubsection{Threshold switching}\label{sec:thresh}
Our thermodynamic approach relates the threshold voltage $U_T$ to the field induced nucleation. \cite{karpov2008,karpov2008a,karpov2008b,nardone2009} Omitting the details, it can be presented, in Gaussian units, as [from e. g. Eq. (13) of Ref. \onlinecite{karpov2008} and with additional multiplier 1/2 derived in Ref. \onlinecite{karpov2015}],
\begin{equation}\label{eq:UT}
U_T=\frac{hW_0}{kT\ln (\tau /\tau _0)}\sqrt{\frac{3\pi ^3\alpha ^3\Lambda W_0}{32\varepsilon r_c^3}}\approx \frac{12h}{kT\sqrt{\varepsilon}}\frac{(\sigma r_{\rm min})^{3/2}}{\ln (\tau /\tau _0)}\end{equation}
where $$W_0=16\pi \sigma^3/\delta\mu ^2 \quad {\rm and}\quad  r_c=2\sigma /\delta \mu$$ are the classical nucleation barrier and radius, \cite{kaschiev2000} $\sigma$ and $$\delta\mu\approx \delta\mu _{1,2} \equiv \mu _1-\mu _2$$ are the interfacial tension and the difference in chemical potentials  between the insulating host and CF, $\alpha =r_{\rm min}/r_c\sim 0.1$, $r_{\rm min}$ is the minimum CF radius (consistent with its integrity), $\varepsilon$ is the dielectric permittivity of the host material, $\tau$ is the electric pulse length, and $\tau _0\sim 10^{-13}$ s is the characteristic atomic vibration time in solids; $\Lambda$ is a multiplier logarithmically dependent on system parameters and not too different from unity.

The first part on the right hand side of Eq. (\ref{eq:UT}) represents the result from Ref. \onlinecite{karpov2008}; it is more  convenient for numerical estimates because the characteristic $W_0\sim 1$ eV and $r_c\sim 1$ nm are well known for solids. The second part presentation shows explicitly that $U_T$ does not depend on $\delta\mu$, which can be either positive or negative describing nucleation of thermodynamically stable or metastable conductive embryo. \cite{nardone2009}

We estimate $\sigma\sim 0.01$ J/m$^2$ based on $r_c\sim 1$ nm and the above mentioned $\delta\mu _1\gtrsim \delta \mu _2\sim 10^9$ J/m$^3$. Setting also $r_{\it min}\sim 0.1$ nm, $h\sim 20$ nm, $\ln (\tau/\tau _0)\sim 10$, $\varepsilon\sim 25$ and $T\sim 600$ K (due to the Joule heat\cite{sharma2015}), yields $U_T\sim 0.6$ V, consistent with the typical data.\cite{fantini2013}

We recall, that the mechanism of field induced nucleation \cite{karpov2008,karpov2008a,karpov2008b,nardone2009} is based on a strong reduction of the electric field energy due to nucleation of a conductive needle shaped embryo. Once created, the field strength is further amplified towards its tip (lightning rod effect). Therefore, nucleation of the next embryos at the tip becomes easier, and the probability of formation of a narrow CF  is determined by the first nucleation event at the threshold voltage given in Eq. (\ref{eq:UT}).
The radial growth of a just formed narrow CF with $r\sim r_{\rm min}$ is described in Sec. \ref{sec:radgr} below.

Note that the above description defines the threshold voltage through the condition
\begin{equation}\label{eq:UT1}\tau =\tau _0\exp\left(\frac{\tilde{U}}{U}\right) \quad {\rm when}\quad  U=U_T\end{equation}
where $\tilde{U}$ is presented by an obvious combination of parameters from Eq. (\ref{eq:UT}), for example,
\begin{equation}\label{eq:Utilde}
\tilde{U}=\frac{hW_0}{kT}\sqrt{\frac{3\pi ^3\alpha ^3\Lambda W_0}{32\varepsilon r_c^3}}.\end{equation}
If the field increases with time, so that $U=\lambda t$ (used in some experimental studies) , then the probability $p$ of nucleation is described by the equation
\begin{equation}\label{eq:dpdt}\frac{dp}{dt}=\frac{1}{\tau _0}\exp\left(-\frac{\tilde{U}}{U}\right).\end{equation}
Integrating the latter and setting $p=1$ defines the threshold voltage through the equation,
\begin{equation}\label{eq:UT2}U_T\approx \tilde{U}\left[\ln\left(\frac{U_T^2}{\lambda\tilde{U}\tau _0}\right)\right]^{-1}, \end{equation}
The transcendental Eq. (\ref{eq:UT2}) can be easily iterated by replacing $U_T$ under the logarithm with its approximate value starting with $U_T=\tilde{U}$, then
\begin{equation}\label{eq:UT3}U_T\approx \tilde{U}\left[\ln\left(\frac{\tilde{U}}{\lambda\tau _0}\right)\right]^{-1}, \end{equation}
etc., where Eq. (\ref{eq:UT3}) provides a rather close approximation with the accuracy of $\sim 10$\%. It predicts that $U_T$ should increase with the sweep rate $\lambda$, consistent with the data.


Another aspect of nucleation switching important for non-crystalline nano-devices is its stochastic nature. It was shown \cite{karpov2008a,karpov2008b} that, because of the inherent disorder, the delay times of switching and the threshold voltages are statistically distributed and the width of these statistical distributions decreases with the area of a structure (i. e. a CF cross section) where the nucleation takes place.  Vive versa, the increase in $U_T$ variations is due to suppression of self-averaging with the area decrease. The underlying physics is that the field induced nucleation in a  RRAM structure takes place through the gap of the preliminary formed filament whose cross sectional area is rather small, on the order of several nanometers. Therefore, that nucleation evolves along the easiest of the available pathways, which in a given filament does not necessarily represent the entire statistical distribution.

A more quantitative analysis of that issue for RRAM devices goes beyond the scope of this manuscript. Here we limit ourselves to pointing out that, based just on the above statements, the variations between the parameters of the nominally identical RRAM structures, should decrease with CF area. The latter prediction is in qualitative agreement with the observations presented in Fig. 4 of Ref. \onlinecite{fantini2012} where variations strongly increase with CF resistance that is inversely proportional to the CF area. This aspect of the `nano-glass' behavior is similar to that discussed in the preceding section for random double well potentials.

\subsubsection{Nucleation of insulating gap }\label{sec:insgap}
As explained in Sec. \ref{sec:proc}, the gap formation is triggered by the unfavorable polarization of a host material developed during the SET process. The gap constituting new phase is energetically favorable providing gain $Al\delta \mu _{\max} $ in free energy where $A$ is the gap crosssectional area and $l$ is its width. Here
\begin{equation}\label{eq:deltamu}
\delta\mu _{\max} \approx\delta\mu _1+\delta \mu _2\end{equation}
corresponds to the transition from the unfavorably polarized CF to the insulating phase.
We consider two possible scenarios: complete rupture of CF, $A=A_0$, and partial CF rupture leaving some neck of crosssectional area $A_0-A$ between the gap edges (Fig. \ref{fig:gap}) where $A_0$ is the crosssectional area of CF before gap formation.

\underline{\it Complete rupture}. We assume first that the electronic processes remain fast enough to adiabatically follow changes in atomic configuration, in particular, the electric current through the stack remaining the same due to the corresponding increase of the local electric field (the alternative case is discussed at the end of this subsection). The gap formation  will then change the free energy by,
\begin{equation}\label{eq:gapen1}
F=-\delta\mu _{\max}A_0l+2A_0\sigma +\frac{E^2}{8\pi}\frac{\rho _i^2}{\rho _c^2}A_0l .
\end{equation}

Here $2\sigma A_0$ is the interfacial energy loss. The last term describes the electric field energy due to the interior field $E_{\rm int}$ that must be by the factor $\rho _i/\rho _c\gg 1$ stronger than $E=U/h$  to maintain the current flow through the stack. $F$ becomes negative when
\begin{equation}\label{gapw}
l\geq l_c=r_c\left(1-\frac{E^2\rho _i^2}{8\pi\delta\mu _{\max}\rho _c^2}\right)^{-1}
\end{equation}
where the classical nucleation radius $r_c=2\sigma /\delta\mu _{max}$. \cite{kaschiev2000}

Once the gap is formed, the current will decrease by the factor of
\begin{equation}\label{eq:ratio1}I_D/I_E=\frac{\rho _i}{\rho _c} \left(1-\frac{E^2\rho _i^2}{8\pi\delta\mu _{\max}\rho _c^2}\right)\end{equation}
where $I_D$ and $I_E$ stand for the currents in points E and D in Fig. \ref{fig:cartoon}.
Assuming the typical $E\sim 10^5$ V/cm and $\delta\mu _{\max}\sim 2\cdot 10^9$ J/m$^3$, one can estimate  $E^2/8\pi\delta\mu _{\max}\sim 10^{-6}$, while the ratio $\rho _i/\rho _c$ is sensitive to material properties and vary between different device recipes. (The voltage will change as well due to redistribution between the load and a just formed gap resistance.)

It follows that (a) the snap forward ratio $I_D/I_E$ depends on the ratio of insulating and conductive phase resistivities varying between different materials, and (b) when the latter ratio is high enough, the filament breakup becomes impossible [$I_D/I_E$ cannot be negative in Eq. (\ref{eq:ratio1})]: CF is stabilized by the electric field.


\underline{\it Partial rupture}. Following Fig. \ref{fig:gap}, the interior field $E_{\rm int}$ must be by the factor $A_0/(A_0-A)$ stronger than $E=U/l$  to provide continuous current flow through the gap. As a result the free energy change accompanying the gap formation becomes
\begin{equation}\label{eq:gapen}
F=-\delta\mu _{\max}Al+2A\sigma +\frac{E^2}{8\pi}\frac{A_0^2}{(A_0-A)^2}Al .
\end{equation}
$F$ is stationary when
\begin{equation}\label{eq:gapA}
A_0-A=A_0\left(E^2/8\pi\delta\mu _{\max}\right)^{1/3}\quad (\ll A_0).\end{equation}
The corresponding energy decrease must be greater than the surface energy loss $2\sigma A_0$. That takes place when $l>r_c$.

We observe that the insulating gap can nucleate with a width $l\gtrsim r_c$ nm leaving a narrow bridging neck.
\begin{figure}[hbt]
\includegraphics[width=0.25\textwidth]{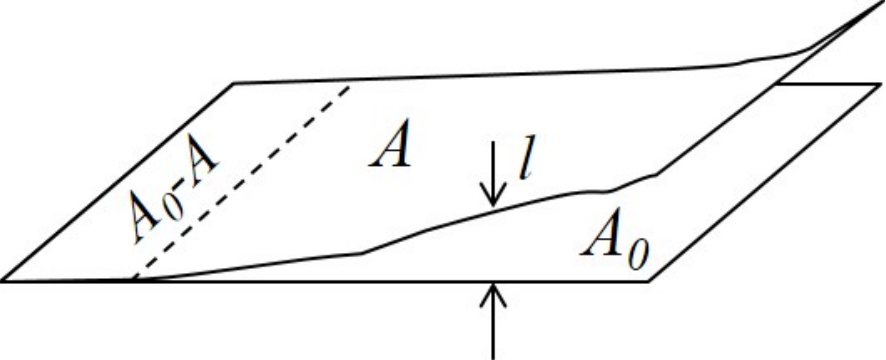}
\caption{Nucleation of an insulating gap of area $A$ and width $l$ in the filament cross-section of area $A_0$. \label{fig:gap}}
\end{figure}
As the gap is formed, the current will snap forward decreasing by the factor of
\begin{equation}\label{eq:ratio}I_D/I_E=A_0/(A_0-A)=(8\pi\delta\mu _{\max}/E^2)^{1/3}\sim 100.\end{equation} The latter prediction is consistent with the data. \cite{fantini2012,wouters2012,wei2015,sawa2005,beck2000}

Comparing free energies in Eqs. (\ref{eq:gapen1}) and (\ref{eq:gapen}) shows that the complete gap rupture is energetically more favorable when $\rho _c/\rho _i<(E^2/8\pi\delta\mu _{\max})^{1/2}\sim 0.01$ where we have used $E\sim 10^6$ V/cm and $\delta\mu _{\max}\sim 10^9$ J/m$^3$. (In the case of very fast structural transition mentioned at the end of preceding subsection, the latter inequality changes to $l/h< 0.01$.)
%

We shall end this section by pointing at its approximations lacking numerical factors and neglecting the concomitant thermal processes that can be significant.\cite{sharma2015} However, this remains the only analytical approach to CF rupture phenomena since their first observations more than 100 years ago (for the contacts of dissimilar metals);\cite{goddard1912}  further efforts are called upon.

\subsection{Growth processes}\label{sec:growth}

Our approach is based on the reduction of a kinetic problem of the filament or gap  growth to the free energy analysis, which we briefly illustrated for the case of CF radii. We start with the kinetic Fokker-Planck equation, which,  for the average CF radius (neglecting variations in an ensemble of different CF) can be transformed to, \cite{karpov2011}
\begin{equation}\label{eq:mobility}
\frac{\partial r}{\partial t}=-b_r\frac{\partial  F}{\partial r}.
\end{equation}
The latter has the standard meaning of a relation between the (growth) velocity and the (thermodynamic) force $-\partial F/\partial r$, with the mobility $b_r$. It follows that the steady  state average radius corresponds to the stationary point of free energy, which condition we use next.

We note that the concept of free energy $F$ is not compromised by the power dissipation, since the electric current is fixed by the circuit and serves only as a temperature source. \cite{karpov2011} The corresponding requirement (of self-consistent Fokker-Planck equation) \cite{selfFP} is that the thermalization time $\tau _T$ must be shorter than that of system evolution, empirically, $\tau _r\sim $ 10-100 ns. Another wording of the same is that the system remains quasistatic with temperature adiabatically following its particular configurations.

For numerical estimates we note that $\tau _T\sim L^2/\kappa$ where $L$ is the characteristic linear dimension of the system and $\kappa$ is the thermal diffusivity.  The latter ratio of thermal conductivity \cite{panzer2009}  $\chi\sim 1$ W/m$\cdot$K over specific heat $c\sim 10$ J/cm$^3\cdot$K is estimated as $\kappa\sim 10^{-3}$ cm$^2$/s. Assuming the nanometer sized devices, $L\sim 1$ nm, yields then $\sim 0.01$ ns. Therefore, the existing RRAM devices fall in the domain $\tau _r\gg \tau _T$ where the thermodynamic analysis applies.

An important particular case represents the thermalization process dominated by CF per se serving as the strongest heat conductor transferring energy to the device electrodes. In that case assumed earlier for the threshold switches \cite{petersen1976} and modern RRAM devices \cite{ielmini2011}
\begin{equation}\label{eq:tauT}
\tau _T=h^2/\kappa
\end{equation}
where $\kappa$ is understood as the thermal diffusivity of Hf based CF. Using the numerical values \cite{ielmini2011} $c\sim 2$ J/cm$^3\cdot$K and $\chi\sim 0.2$ W/m$\cdot$K, it is estimated as $\kappa\sim 0.1$ cm$^2$/s leading to $\tau _T\sim 10^{-12}$ s for a 20 nm long CF, close to the estimate from the preceding paragraph. The difference is that Eq. (\ref{eq:tauT}) predicts the CF length dependent $\tau _T$, which will result in a rather specific prediction of $V_{\rm SET}$ and CF resistance $R$ independent of $h$ given in Sec. \ref{sec:radgr} below.

The major part of the free energy is given by,
\begin{equation}\label{eq:freeen}
F=\int d^3r c\delta T+2\pi rh\sigma +\pi r^2h\delta\mu+\int d^3r\frac{E^2\varepsilon}{8\pi}.
\end{equation}
Here, $c$ is the specific heat, $\delta T$ is the temperature change.  The first term in Eq. (\ref{eq:freeen}) represents the thermal contribution, the second and third correspond to the phase transformation, and the fourth term stands for the electrostatic energy. We approximate the first terms with $\tau _TP$ where $P$ is the Joule power produced by the filament.

\subsubsection{Radial growth of CF}\label{sec:radgr}

The domain BC in Fig. \ref{fig:cartoon} corresponds to the current source regime because the filament dynamic resistance $R\ll R_L$.  The electrostatic energy does not change in the course of filament radius growth and is neglected in what follows. Neglecting also the surface tension term (see the discussion at the end of this subsection), the corresponding free energy can be written as,
\begin{equation}\label{eq:freeen1}
F=\tau _T I^2R+\frac{\rho _c h^2\delta \mu _1}{R}
\end{equation}
where $\rho _c$ is the resistivity of CF phase and we used $R=\rho _ch/\pi r^2$.
Optimizing the latter with respect to $R$ yields its optimum value and the corresponding CF radius,
\begin{equation}\label{eq:param1}
R^{(0)} \equiv\sqrt{\frac{\rho _c h^2\delta \mu _1}{\tau _T I^2}},\quad   r^{(0)} \equiv\left(\frac{\rho _c\tau _T}{\delta \mu _1}\right)^{1/4}\sqrt{\frac{I}{\pi}}.
\end{equation}

These results define the steady state CF resistance and radius. The corresponding SET voltage is given by
\begin{equation}\label{eq:USET}
U_{\rm SET}=R^{(0)}I=U_{\rm SET}^{(0)} \equiv h\sqrt{\frac{\rho _c\delta\mu _1}{\tau _T}}.\end{equation}

A particular important case of CF dominated thermalization in Eq. (\ref{eq:tauT}) makes the CF resistance and SET voltage independent of device thickness,
\begin{equation}\label{eq:USET1}R^{(0)}=\sqrt{\frac{\rho _c \kappa\delta \mu _1}{I^2}}, \quad  r^{(0)}=\left(\frac{\rho _ch^2}{\kappa \delta \mu _1}\right)^{1/4}\sqrt{\frac{I}{\pi}},
\end{equation}
and
\begin{equation}\label{eq:USET2}
U_{\rm SET}^{(0)} =\sqrt{\kappa\rho _c\delta\mu _1}.\end{equation}

Eqs. (\ref{eq:param1}) and (\ref{eq:USET1}) predict the dependence $R\propto I^{-1}$ explaining the observations in Table \ref{tab:questions}. Also, using the above estimated $\delta \mu _1\sim 10^9$ J/m$^3$ and resistivity \cite{ielmini2011} $\rho _c\sim 10^{-4}$ $\Omega\cdot$cm, it predicts the numerical values $R\approx 1$ K$\Omega$, $r\approx 3$ nm, and $U_{\rm SET}\sim 0.1$ V consistent with the data.\cite{wouters2012}

Remarkably, Eqs. (\ref{eq:param1}) and (\ref{eq:USET1}) predict CF resistance that is independent of device thickness $h$. The experimentally established fact of that independence therefore does not require a constriction described in the `hour glass' model \cite{degrave2014,kim2010} mentioned in the beginning of Sec. \ref{sec:mod} above. [We should note however that while Eqs. (\ref{eq:param1}) and (\ref{eq:USET1}) show that the thickness independent CF resistance can be understood without the assumptions about its determining constriction, they do not state that CF constrictions, such as observed, \cite{yalon2015c} cannot exist.]

We note that the phenomenon $R\propto I^{-1}$ experimentally is not limited to RRAM and threshold switch devices: it was observed for 1D granular metals \cite{falcon2004,bequin2010} and, more than 100 years ago, for the granular media (metal filings) forming the coherer devices. \cite{guthe1901} The approach presented here may be relevant for the latter two phenomena as well.

To make this subsection analysis more accurate, one can account for the above neglected surface tension term as a perturbation. It can be conveniently estimated as $2\pi\sigma rh=\pi r^2h\delta\mu (r_c/r)$ where $r_c =2\sigma /\delta\mu $ is the classical nucleation radius \cite{kaschiev2000} whose typical value in solids is of the order of 1 nm. The latter estimate shows that the surface term becomes significant when the filament radius remains small, $r\sim r_c$, but it can be neglected for the `grown' filament with $r\gg r_c$, which empirically corresponds to the vertical portion of the B-C domain.

Adding the surface contribution to the free energy of Eq. (\ref{eq:freeen1}) and optimizing it, to the accuracy of terms linear in $\sigma$, yields,
\begin{equation}\label{eq:corr1} R=R^{(0)}\left[1+\frac{r_c}{4r^{(0)}}\right],  \quad U=U_{\rm SET}^{(0)}\left[1+\frac{r_c}{4r^{(0)}}\right].\end{equation}
Taking into account that $r\propto \sqrt{I}$, we observe that the current voltage characteristic becomes slightly `back-slashed' (i. e. showing some negative slope), in qualitative agreement with the available data.

Finally we note that combining Eqs. (\ref{eq:USET1}), (\ref{eq:USET2}), and (\ref{eq:rate}) with $N_a=\pi r^2h/a_0^3$, yields the relation between the SET process time $\tau$ and its driving current $I$ (representing here the compliance current, i. e. the maximum current on the domain B-C of Fig. \ref{fig:cartoon} allowed by the setup),
\begin{equation}\label{eq:tauI}
\ln\left(\frac{\tau}{\tau _0}\right)=\frac{W_a}{\delta\mu _1a_0^3+kT_0}.\end{equation}
The first term in the denominator describes the effect of temperature increase $k\delta T$ and turns out to be independent of $I$. Its physical interpretation is that the Joule heat generated thermal energy increase must be equal the chemical energy  in order to overcome the energy deficit $\delta\mu _1$ per volume in Fig. \ref{fig:CP}. Another useful form of the latter result concerns the freezing temperature,
\begin{equation}\label{eq:Tf}
T_f= T_0+\frac{a_0^3\delta\mu _1}{k}=\frac{W_a}{\ln(\tau /\tau _0)}\end{equation}
and emphasize its thermodynamic nature.

Using the above numerical values $a_0=0.2$ nm and $\delta \mu _1=10^9$ J/m$^3$ yields $\delta T\sim 600$ K, which is consistent with the data. \cite{yalon2015a} However, the applicability of Eq. (\ref{eq:tauI})  is limited to not very high $\delta T$ (practically well below 1000 K) allowing the above thermal analysis without radiation cooling.

As a final note, we mention that $\tau$ in Eq. (\ref{eq:tauI}) has the meaning of the characteristic time of radial filament expansion, which exponentially decreases with the temperature $T_f$ (and thus heat $\delta\mu _1$ per volume) necessary to maintain that process. Also, it should be remembered that Eqs. (\ref{eq:tauI}) and (\ref{eq:Tf}) are limited to the case of endothermic reaction (i. e. $\delta\mu _1>0$) when the relation $r\propto \sqrt{I}$ applies; it cannot be extended to the alternative case of $\delta\mu _1>0$.

\subsubsection{Growth of insulating gap}\label{sec:gapgr}

Consider the opposite regime of voltage source operations ($R\gg R_L$) corresponding to the RESET domain E-F in Fig. \ref{fig:cartoon}. As illustrated in Fig. \ref{fig:CP}, it is characterized by the change in chemical potential, $\delta\mu '=\mu _{i}-\mu _{mc}=-\delta\mu _1+\delta\mu _2<0$ when the insulating gap is formed as a final product of the structural transformation involved. However, as explained in Sec. \ref{sec:proc} above and illustrated in  Fig. \ref{fig:CP} there is a significant difference in the transformation barriers, suggesting that the insulating gap is formed through the intermediate unstable state requiring increase $\delta \mu _2>0$ in chemical potentials. After the energy $\delta\mu _2\pi r^2l$ is provided, the unstable CF quickly decays into the stable insulating phase. The free energy responsible for the former bottleneck process is described by,
\begin{equation}\label{eq:freeen2}
F=\frac{\tau _T U^2}{R_i}+\delta\mu _2\pi r^2+\frac{E^2\varepsilon}{8\pi}\pi r^2l
\end{equation}
where $l$ stands for the gap width, $R_i=\rho _il/\pi r^2$ and $\rho _i$ represent its resistance and resistivity.

The behavior of the electrostatic term in Eq. (\ref{eq:freeen2}) depends on the relation between the gap growth time $t_g$ and the characteristic $RC$ time of the system. Assuming $RC\ll t_g$ the system remains in equilibrium with the voltage source; hence voltage $U$ is given, and the field strength becomes $U/l$ yielding the electrostatic term inversely proportional to $l$, similar to the first term in Eq.  (\ref{eq:freeen2}). The electrostatic contribution decreases with $l$, because maintaining constant voltage results in passing a charge through the voltage source. \cite{landau1984} (In particular, $CU^2\delta l/2l$ is the energy gain due to increase $\delta l$ in the distance $l$ between the plates of a parallel plate capacitor $C$ at a fixed voltage $U$.) Based on the experimental values \cite{fantini2012} we assume here that $RC\ll t_g$.


With the above in mind, minimizing the free energy in Eq. (\ref{eq:freeen2}) leads to the equation
\begin{equation}\label{eq:equil}
-\frac{\tau _TU^2\pi r^2}{\rho _il^2}+\delta \mu _2\pi r^2+\frac{\varepsilon U^2 r^2}{8l^2}=0.\end{equation}
Here the first and third terms have similar $l$-dependencies, and the latter one is small for any practical choice of material parameters, for example, $\tau _T\sim 0.01$ ns, $\varepsilon \approx 25$ (for \cite{huang2010} HfO$_2$) and  $\rho _i\sim 0.001-0.1$ $\Omega\cdot$m.

Solving Eq. (\ref{eq:equil}) yields the gap width $l$, its resistance $R_i$ and the current $I$ that should be identified with the `saturation' current, $I_{\rm R,SAT}$ marked in Fig. \ref{fig:cartoon} domain E-F,
\begin{equation}\label{eq:param2}
l=\sqrt{\frac{\tau _T U^2}{\rho _i\delta\mu _2}} \quad {\rm and} \quad I_{\rm R,SAT}=\frac{U}{R_i}=r^2\sqrt{\frac{\delta\mu _2}{\tau _T\rho _i}}.
\end{equation}

For numerical estimates, we assume $r\sim 10$ nm and $\rho _i\sim 100\rho _c\sim 10^{-2}$ $\Omega\cdot$m based on the typical difference in the ON and OFF state resistances.\cite{hildebrandt2011}  This yields a reasonable gap $l\sim 1$ nm and $I_{\rm R,SAT}\sim 10$ $\mu$A according to Eq. (\ref{eq:param2}) in fair agreement with the data. \cite{fantini2012,lee2010,kalantarian2012} The sometime observed deviations from the voltage independent current in the domain E-F of Fig. \ref{fig:cartoon} can be caused by the non-ohmicity of the insulating phase resistivity.

Taking into account the discussion at the end of Sec. \ref{sec:proc}, Eq. (\ref{eq:param2}) predicts the `saturation' current, $I_{\rm R,SAT}$ being proportional to $\delta\mu _2$ should be time dependent. Such dependencies have been observed. For example, in the experimental design of Ref. \onlinecite{fantini2012}, the time $t$ that must be substituted in Eq. (\ref{eq:log}) is determined by the change in the electric potential divided by the voltage ramp rate $|dV/dt|$ leading to the observed dependence  $I_{\rm R,SAT}$ vs. $|dV/dt|$. 

For completeness, we will point at an alternative RESET scenario where the domains E-F and F-0 overlap without hysteresis. One can consider indeed that, in spite of a certain increase in $U^2$, the increase in resistance at the E-F domain suppresses Joule heat enough to ensure that the freezing criterion in Eq. (\ref{eq:rate}) obeys.  Should that condition take place, the system would not structurally evolve in the domain E-F resulting in the no hysteresis behavior, and  Eq. (\ref{eq:param2}) becomes unapplicable.

Finally, we note that our above phenomenological theory is limited to the ohmic mechanism of conductivity setting aside possibilities of electron tunneling \cite{long2013,lv2015,li2015} that would change the results in Eq. (\ref{eq:param2}). Therefore, we we would like to briefly describe the effects of quantum tunneling through the gap dielectric.

In our generic approach we use the simplest expression $R_T=R_T^0\exp(l/a_T)$ for the tunneling resistance $R_T$ vs. gap width $l$ where $R_T^0$ and $a_T$ are two phenomenological parameters. Using $R_T$ instead of $R_i$ and optimizing the free energy in Eq. (\ref{eq:freeen2}) yields,
\begin{equation}\label{eq:quant}
l=a_T\ln\left(\frac{\tau _TU^2}{R_T^0a_T\delta\mu _2\pi r^2}\right)\quad {\rm and}\quad I=\frac{a_T\delta\mu _2\pi r^2}{\tau _T|U|}.\end{equation}
We conclude that the gap logarithmically widens and tunneling current decreases as $1/U$ with voltage increase.

For numerical estimates we assume $R_T^0\sim 10$ k$\Omega$ (of the order of the quantum resistance \cite{long2013,lv2015,li2015}), $a_T\sim 1$ nm (typical of tunneling in solids), $r\sim 5$ nm, and the above introduced $\tau _T\sim 0.01$ ns, $\delta\mu _2\sim 10^9$ J/m$^3$. With the latter numbers, Eq. (\ref{eq:quant}) yields $l\sim a_T\sim 1$  nm and $I\sim 10$ $\mu$A for $|U|\sim 1$ V. It is worth noting that the latter quantum current is in the order of magnitude equal $I_{\rm R,SAT}$.

Because the tunneling contribution decreases, the ohmic current will dominate starting from some voltage. A simple extrapolation of such a behavior takes the form
\begin{equation}\label{eq:fit}
I=P1+\frac{P2}{U}\end{equation}
where $P1$ and $P2$ are two parameters that can be determined from experiments and which characteristic values are provided respectively in Eqs. (\ref{eq:param2}) and (\ref{eq:quant}).

\section{Conclusions}\label{sec:concl}

We have derived closed form equations for all the quantities in question listed in Table \ref{tab:questions}.

Our results are summarized in the related Table \ref{tab:answers}. The corresponding numerical estimates, while approximate, fall in the ballpark of measured values.

The essence of our phenomenological theory is (a) the notion of the filament charging, (b) its accompanying polarization of the host material, and (c) the existence of three phase states of the material: stable insulating, unstable conducting, and long-lived metastable, conducting. The items (a) and (b) are model independent, while (c) remains a  model hypotheses, which however suffice to explain a large number of outstanding questions as illustrated in Tables \ref{tab:questions} and \ref{tab:answers}.
\begin{table}[h]
\caption{Answering the questions of Table \ref{tab:questions}. }
  \begin{tabular}{|p{4cm} | p{4cm} |}
    \hline
   Domain & Answer \\ \hline
   A-B   & Threshold voltage: Eq. (\ref{eq:UT}).  \\  \hline
   B-C & SET voltage: Eq. (\ref{eq:USET}).  \\  \hline
   C-0-D& Filament radius and resistance: Eq. (\ref{eq:param1}) \\  \hline
   D-E & $U_{\rm RESET}=-U_{\rm SET}$ and $I_{\rm RESET}=-I_{\rm SET}$, Eq. (\ref{eq:ratio}).\\  \hline
   E-F & Expressions for $I_{\rm R,SAT}$: Eq. (\ref{eq:param2})  \\  \hline
  \end{tabular}\label{tab:answers}
  \end{table}


Finally, our results contain a number of predictions calling upon experimental verification. Such is the phenomenon of filament charging, the temperature dependence of threshold voltage, the amplitude of the current snap-forward, voltage dependence of insulating gap width and some others.

\section*{Acknowledgement}
This work was supported in part by the Semiconductor
Research Corporation (SRC) under Contract No. 2016-LM-2654.


\begin{thebibliography}{99}


\bibitem{slonczewski1996}J. C. Slonczewski, Current-driven excitation of magnetic multilayers, Journal of Magnetism and Magnetic Materials, {\bf 159}, L1 (1996).
\bibitem{fantini2012}A. Fantini, D. J. Wouters, R. Degraeve, L. Goux, L. Pantisano, G. Kar, Y. -Y. Chen, B. Govoreanu, J. A. Kittl, L. Altimime, M. Jurczak, Intrinsic Switching Behavior in HfO2 RRAM by Fast Electrical Measurements on Novel 2R Test Structures, 2012 4th IEEE International Memory Workshop, Milan, May 20-23 (2012), IEEE, DOI:10.1109/IMW.2012.6213646
\bibitem{wouters2012}D.J. Wouters, L. Zhang, A. Fantini, R. Degraeve, L. Goux, Y.Y. Chen, B. Govoreanu, G.S. Kar, G. V. Groeseneken, and, M. Jurczak, Analysis of Complementary RRAM Switching, IEEE Electron. Dev. Lett., {\bf 33}, 1186 (2012).
\bibitem{wei2015}Z. Wei, K. Eriguchi, S. Muraoka, K. Katayama, R. Yasuhara, K. Kawai, Y. Ikeda, M. Yoshimura, Y. Hayakawa, K. Shimakawa, T. Mikawa, and S. Yoneda, Distribution Projecting the Reliability for 40 nm ReRAM and beyond based on Stochastic Differential Equation, 2015 IEEE International Electron Devices Meeting (IEDM), Washington, DC, Dec 7-9 (2015), DOI: 10.1109/IEDM.2015.7409650
\bibitem{sawa2005}A. Sawa, T. Fujii, M. Kawasaki, and Y. Tokura, Colossal Electro-Resistance Memory Effect at Metal/La2CuO4 Interfaces, Jpn. J. Appl. Phys. {\bf 44}, L1241 (2005).
\bibitem{beck2000}A. Beck, J. G. Bednorz, Ch. Gerber, C. Rossel, and D. Widmer, Reproducible switching effect in thin oxide films for memory applications, Appl. Phys. Lett. {\bf 77}, 139 (2000).
\bibitem{degrave2014}R. Degraeve, A. Fantini, N. Raghavan, L. Goux, S. Clima, Y. Y. Chen, A. Belmonte, S. Cosemans, B. Govoreanu, D. J. Wouters, Ph. Roussel, G. S. Kar, G. Groeseneken, M. Jurczak, Hourglass concept for RRAM: A dynamic and statistical device model, Proceedings of the 21th International Symposium on the Physical and Failure Analysis of Integrated Circuits (IPFA), 245 (2014), DOI: 10.1109/IPFA.2014.6898205
\bibitem{kim2010}K. M. Kim, M. H. Lee, G. H. Kim, S. J. Song, J. Y. Seok, J. H. Yoon, and C. S. Hwang, Understanding structure-property relationship of resistive switching oxide thin films
using a conical filament model, Appl. Phys. Lett. {\bf 97}, 162912 (2010).
\bibitem{celano2014}U. Celano, L. Goux, A. Belmonte, K. Opsomer, A. Franquet, A. Schulze, C. Detavernier, O Richard, H. Bender, M. Jurczak, and W. Vandervorst, Three-Dimensional Observation of the Conductive Filament in
Nanoscaled Resistive Memory Devices, Nano Lett.,  {\bf 14}, 2401 (2014).
\bibitem{privetera2013}S. Privitera, G. Bersuker, B. Butcher, A. Kalantarian, S. Lombardo, C. Bongiorno,
R. Geer, D.C. Gilmer, P.D. Kirsch, Microscopy study of the conductive filament in HfO2 resistive switching
memory devices, Microelectronic Engineering {\bf 109}, 75 (2013).
\bibitem{halliday}D. Halliday, R. Resnik, and J. Walker, {\it Fundamentals of Physics}, 10th edition, Wiley (2014).
\bibitem{muller2015}J. Muller, P. Polakowski S. Mueller, and T. Mikolajick, Ferroelectric Hafnium Oxide Based Materials and Devices,
Assessment of Current Status and Future Prospects, ECS Journal of Solid State Science and Technology, {\bf 4} N30 (2015).
\bibitem{xu1999} Y. Xu, J.D. Mackenzie, A theoretical explanation for ferroelectric-like properties of
amorphous Pb(Zr$_x$Ti$_{1-x}$O$_3$ and BaTiO$_3$, Journal of Non-Crystalline Solids, {\bf 246}, 136 (1999).
\bibitem{zhao2002}X. Zhao, D. Vanderbilt, First-principles study of structural, vibrational, and lattice dielectric properties of hafnium oxide, Phys. Rev. B, {\bf 65}, 233106 (2002).
\bibitem{cooray2007}M. L. C. Cooray and V. G. Karpov, Long range fluctuations in thin-film structures, Phys. Rev. B. {\bf 75}, 155303 (2007).
\bibitem{maxwell}J.C. Maxwell, On the Electrical Capacity of a long narrow Cylinder, and of a Disk of sensible Thickness, in {\it Scientific Papers} (Dover, New York, 1965), Vol. II, p. 672.
\bibitem{mcdonald2003}K. T. McDonald, Capacitance of a Thin Conducting Disk and of Conducting Spheroids, available at \url{http://physics.princeton.edu/~mcdonald/examples/index.html} .
\bibitem{scharstein2007}R. W. Scharstein, Capacitance of a tube, Journal of Electrostatics {\bf 65}, 21, (2007).
\bibitem{moynihan1974}C. T. Moynihan, A. J. Easteal, J. Wilder, Dependence of the Glass Transition Temperature on Heating and Cooling Rate, J. Phys. Chem., {\bf 78}, 2673, (1974).
\bibitem{ald1990}S. B Aldabergenova, N. A Feoktistov, V. G Karpov, K. V Koughia, A. B Pevtsov, V. U Solovijev, Thermally induced metastable processes in amorphous hydrogeneted silicon, in {\it Transport, Correlation and Structural Defects}, p. 129, Edited by: H Fritzsche, Advances in Disordered Semiconductors: Volume 3, World Scientific (1990).
\bibitem{street1991}R. A. Street, {\it Hydrogenated Amorphous Silicon}, Cambridge University Press, New York, Sydney, 1991.
\bibitem{kruchihin2015}V.N. Kruchinin, V.Sh. Aliev, T.V. Perevalov, D.R. Islamov, V.A. Gritsenko, I.P. Prosvirin,
C.H. Cheng, A. Chin, Nanoscale potential fluctuation in non-stoichiometric HfOx and low
resistive transport in RRAM, Microelectronic Engineering {\bf 147}, 165 (2015).
\bibitem{ielmini2011}D. Ielmini, Modeling the Universal Set/Reset Characteristics of Bipolar RRAM by Field- and Temperature-Driven Filament Growth, IEEE Trans. Electron Devices, {\bf 58}, 4309 (2011).
\bibitem{coiocchini2016}N. Ciocchini, M. Laudato, M. Boniardi, E. Varesi, P. Fantini, A. L. Lacaita and D. Ielmini, Bipolar switching in chalcogenide phase change memory, Scientific Reports, {\bf 6}, 29162 (2016).

\bibitem{maestro2015}M. Maestro, J. M.-Martinez, J. Diaz, A.C.-Yepes, M.B. Gonzalez, R. Rodriguez, F. Campabadal, M. Nafria, and X. Aymerich, Analysis of Set and Reset mechanism in Ni/HfO2-based RRAM with fast ramped voltages, Microelectronic Engineering, {\bf 147},176 (2015).
\bibitem{schindler2009}C. Schindler, G. Staikov, and R. Waser, Electrode kinetics of Cu-SiO2-based resistive switching cells: Overcoming the voltage-time dilemma of electrochemical metallization memories, Appl. Phys. Lett. {\bf 94}, 072109 (2009).
\bibitem{yalon2015a}E. Yalon, A. A. Sharma, M. Skowronski, J. A. Bain, D. Ritter, I. V. Karpov, Thermometry of Filamentary RRAM Devices, IEEE Trans. Electron Devices, {\bf 62}, 2972 (2015).
\bibitem{erdemir2009}D. Erdemir, A. Y. Lee, and A. S. Meyerson, Nucleation of Crystals from Solution: Classical
and Two-Step Models, Accounts of Chemical Research, {\bf 42}, 621, (2009).
\bibitem{govoreanu2013}B. Govoreanu, A. Ajaykumar, H. Lipowicz, Y.Y. Chen, J.C. Liu, R. Degraeve, L. Zhang, S. Clima, L. Goux, I.P. Radu, A. Fantini, N. Raghavan, G.S. Kar, W. Kim, A. Redolfi, D.J. Wouters, L. Altimime and M. Jurczak, Performance and reliability of Ultra-Thin HfO2-based RRAM (UTO-RRAM), 2013 5th IEEE International Memory Workshop, Monterey, CA, 2013, pp. 48-51.
doi: 10.1109/IMW.2013.6582095
\bibitem{park2016}J. Park, E. Cha, I. Karpov, and H. Hwang, Dynamics of electroforming and electrically driven insulator-metal transition
in NbOx selector, Appl. Phyys. Lett., {\bf 108}, 232101 (2016)

\bibitem{landau1980}L. D. Landau and E. M. Lifsitz, {\it Statistical Physics}, Pergamon, Oxford, (1980).
\bibitem{yalon2015}E. Yalon, A. Gavrilov, S. Cohen, and D. Ritter, Validation and Extension of Local Temperature
Evaluation of Conductive Filaments in RRAM Devices, IEEE Transactions on Electron Devices, {\bf 62}, 3671 (2015).
\bibitem{landau1984}L. D. Landau and E. M. Lifsitz, {\it Electrodynamics of Continuous Media}, Pergamon, Oxford, (1984).
\bibitem{meister2011}S. Meister, S-B Kim, J. J. Cha, H.-S. P. Wong, and Y. Cui, In Situ Transmission Electron
Microscopy Observation of Nanostructural Changes in Phase-Change Memory, ACS Nano, {\bf 5}, 2742 (2011).
\bibitem{karpov2007} I. V. Karpov, M. Mitra, D. Kau, and G. Spadini, Y. A. Kryukov and V. G. Karpov, Fundamental drift of parameters in chalcogenide phase change memory, J. Appl. Phys., {\bf 102}, 124503 (2007).
\bibitem{kogan1996}Sh. Kogan, {\it Electronic noise and fluctuations in solids}, Cambridge University Press, 1996.
\bibitem{hooge1997}F.N. Hooge, P.A. Bobbert, On the correlation function of 1/f noise, Physica B, {\bf 239}, 223 (1997).
\bibitem{fantini2015}A. Fantini, G. Gorine1,, R. Degraeve, L. Goux, C.Y. Chen, A. Redolfi1 S. Clima, A. Cabrini, G. Torelli, M. Jurczak, Intrinsic Program Instability in HfO2 RRAM and consequences on program algorithms,  Electron Devices Meeting (IEDM), 7-9 Dec. 2015, IEEE International, DOI: 10.1109/IEDM.2015.7409648 .


\bibitem{maestro2016}M. Maestro, J. Díaz, A. Crespo-Yepes, M. B. González, J. Martín-Martínez, R. Rodríguez, M. Nafría, F. Campabadal, X. Aymerich, New high resolution Random Telegraph Noise (RTN) characterization method for resistive RAM, Solid State Electronics, {\bf 115}, 140 (2106).
\bibitem{kirton1989}M.J. Kirton, M.J. Uren, Noise in solid-state microstructures: A new perspective on individual defects, interface states and low-frequency 1/f noise, Advances in
Physics, {\bf 38}, 367 (1989).
\bibitem{tsai2013}Shih-Chang Tsai, San-Lein Wu, Jone-Fang Chen, Kai-Shiang Tsai, Tsung-Hsien Kao, Chih-Wei Yang, Cheng-Guo Chen, Kun-Yuan Lo, Osbert Cheng, Yean-Kuen Fang and Shoou-Jinn Chang, Correlation between 1/f Noise Parameters and Random Telegraph Noise in 28-nm High-k/Metal Gate pMOSFETs with Embedded SiGe Source/Drain, Extended Abstracts of the 2013 International Conference on Solid State Devices and Materials, Fukuoka, 2013,PS-3-4 p.62
\bibitem{karpov1992}V. G. Karpov, Fluctuations in the thermal expansion of disordered systems, Pis'ma Zh.Eksper.Teor.Fiz. {\bf 55}, 59 (1992) [Sov.Phys. JETP Letters {\bf 55}, 60 (1992)].



\bibitem{chen2016}C. Y. Chen, A. Fantini, L. Goux, G. Gorine, A. Redolfi, G. Groeseneken, and M. Jurczak, Novel Flexible and Cost-Effective Retention
Assessment Method for TMO-Based RRAM, IEEE Electron Device Lett., {\bf 37}, 1112 (2016).





\bibitem{karpov2008}V. G. Karpov, Y. A. Kryukov, I. V. Karpov, and M. Mitra, Field induced nucleation in glasses, Phys. Rev. B {\bf 78}, 052201 (2008).
\bibitem{karpov2008a}I. V. Karpov, M. Mitra, D. Kau, G. Spadini, Y. A. Kryukov, and V. G. Karpov, Evidence of field induced nucleation in phase change memory, Appl. Phys. Lett. {\bf 92}, 173501 (2008).
\bibitem{karpov2008b}V. G. Karpov, Y. A. Kryukov, I. V. Karpov, and M. Mitra, Crystal nucleation in phase change memory, J. Appl. Phys. {\bf 104}, 054507 (2008).
\bibitem{nardone2009}M. Nardone, V. G. Karpov, C. Jackson, and I. V. Karpov, Unified Model of Nucleation Switching, Appl. Phys. Lett. {\bf 94}, 103509 (2009).
\bibitem{karpov2015}V. G. Karpov, R. E. E. Maltby, I. V. Karpov, and E. Yalon, Phys. Rev. Appl., {\bf 3}, 044004 (2015).
\bibitem{kaschiev2000}D. Kaschiev, {\it Nucleation: Basic Theory with Applications} (Butterworth-Heinemann, Oxford, Amsterdam, 2000).
\bibitem{fantini2013}A.Fantini, L. Goux, R. Degraeve, D.J. Wouters. N. Raghavan, G. Kar, A. Belmonte, Y.Y. Chen, B. Govoreanu, and M. Jurczak, Intrinsic Switching Variability in HfO2 RRAM, IEEE International Memory Workshop, p. 30 (2013). DOI: 10.1109/IMW.2013.6582090.
\bibitem{sharma2015}A. A. Sharma, I. V. Karpov, R. Kotlyar, J. Kwon, Dynamics of electroforming in binary metal oxide-based resistive switching
memory, J. Appl. Phys. {\bf 118}, 114903 (2015).

\bibitem{goddard1912}R. H. Goddard, On the conduction of electricity at contacts of dissimilar metals, Phys. Rev., {\bf 34}, 423 (1912).

\bibitem{karpov2011}V. G. Karpov, M. Nardone, and M. Simon, Thermodynamics of second phase conductive filaments, J. Appl. Phys., {\bf 109}, 114507 (2011)
\bibitem{selfFP}S. Nayakshin and F. Mela, Self-consistent Fokker-Planck treatment of particle distributions in astrophysical plasmas, The Astrophysical Journal Supplement Series, {\bf 114}, 269 (1998).

\bibitem{panzer2009}M. A. Panzer, M. Shandalov, J. A. Rowlette, Y. Oshima, Y. W. Chen, P. C. McIntyre, and K. E. Goodson, Thermal Properties of Ultrathin Hafnium
Oxide Gate Dielectric Films, IEEE Electron Device Letters, {\bf 30}, 1269 (2009).
\bibitem{petersen1976} K. E. Petersen and D. Adler, State of Amorphous Threshold Switches, J. Appl. Phys. {\bf 47}, 256 (1976).

\bibitem{falcon2004}E. Falcon, B. Castaing, and M. Creyssels, Nonlinear electrical conductivity in a 1D granular medium, Eur. Phys. J. {\bf B 38}, 475 (2004).
\bibitem{bequin2010}P. Béquin and V. Tournat, Electrical conduction and Joule effect in one-dimensional chains
of metallic beads: hysteresis under cycling DC currents and influence of electromagnetic pulses, Granular Matter, DOI 10.1007/s10035-010-0185-8 (2010).
\bibitem{guthe1901}K.E. Guthe, On the action of the coherer. Phys. Rev. {\bf 12}, 245 (1901). K.E. Guthe, A. Trowbridge, On the theory of the coherer, Phys. Rev. {\bf 11}, 22 (1900).
\bibitem{huang2010} A. P. Huang, Z. C. Yang, and P. K. Chu, Hafnium-based High-k Gate Dielectrics, in {\it Advances in Solid State Circuits Technologies}, p. 333, Edited by: Paul K. Chu,
 ISBN 978-953-307-086-5, (2010) INTECH, Croatia.







\bibitem{yalon2015c}E. Yalon E, I. Karpov, V. Karpov, I. Ries, D. Kalaev, D. T. Ritter, Detection of the insulating gap and conductive filament growth direction in resistive memories, Nanoscale, {\bf 7}, (2015).
\bibitem{hildebrandt2011}E. Hildebrandt, J. Kurian, M.M. Muller, T. Schroeder, H.-J. Kleebe, and L. Alff, Controlled oxygen vacancy p-type conductivity in HfO2-x thin films, Appl. Phys. Lett. {\bf 99},112902 (2011).

\bibitem{lee2010}H.Y. Lee, Y.S. Chen, P.S. Chen, T.Y. Wu, F. Chen, C.C. Wang, P.J. Tzeng, M.-J. Tsai, and C. Lien, Low-Power and Nanosecond Switching in Robust Hafnium Oxide Resistive Memory With a Thin Ti Cap, IEEE Electron. Dev. Lett., {\bf 31}, 44 (2010).
\bibitem{kalantarian2012}A. Kalantarian, G. Bersuker, D.C. Gilmer, D. Veksler, B. Butcher, A. Padovani, O. Pirrotta, L. Larcher, R. Geer, Y. Nishi, and P. Kirsch, Controlling Uniformity of RRAM Characteristics Through the Forming Process, IEEE international reliability physics symposium (IPRS), 15-19 April 2012, Anaheim CA. p. 6C.4.1-4.5. (2012)


\bibitem{long2013}S. Long, X. Lian, C. Cagli, X. Cartoixà, R. Rurali, E. Miranda, D. Jiménez, L.Perniola, M. Liu and J. Suñé, Quantum-size effects in hafnium-oxide resistive switching, Appl. Phys. Lett., {\bf 102}, 183505 (2013).


\bibitem{lv2015}H. Lv, X. Xu, P. Sun, H. Liu, Q. Luo, Q. Liu, W. Banerjee, H. Sun, S. Long, L. Li and M. Liu,
Atomic View of Filament Growth in Electrochemical Memristive Elements, Scientific Reports 5, Article number: 13311 (2015)
doi:10.1038/srep13311

\bibitem{li2015}Y. Li, S. Long, Y. Liu, C. Hu, J. Teng, Q. Liu, H. Lv, J. Suñé and M. Liu, Conductance Quantization in Resistive Random Access Memory, Nanoscale Research Letters {\bf 10}, 420 (2015), DOI: 10.1186/s11671-015-1118-6

%
%


%
%
%

\end{thebibliography}
\end{document}